# DESIGN AND FABRICATION OF A MICRO ELECTROSTATIC VIBRATION-TO-ELECTRICITY ENERGY CONVERTER

*Yi Chiu\*, Chiung-Ting Kuo and Yu-Shan Chu*

Department of Electrical and Control Engineering, National Chiao Tung University
1001 Ta Hsueh Road, Hsinchu 300, Taiwan, R.O.C.
Tel: +886-3-573-1838, Fax: +886-3-571-5998, Email: yichiu@mail.nctu.edu.tw

**ABSTRACT**

This paper presents a micro electrostatic vibration-to-electricity energy converter. For the 3.3 V supply voltage and 1cm$^2$ chip area constraints, optimal design parameters were found from theoretical calculation and Simulink simulation. In the current design, the output power is 200 µW/cm$^2$ for the optimal load of 8 MΩ. The device was fabricated in a silicon-on-insulator (SOI) wafer. Mechanical and electrical measurements were conducted. Residual particles caused shortage of the variable capacitor and the output power could not be measured. Device design and fabrication processes are being refined.

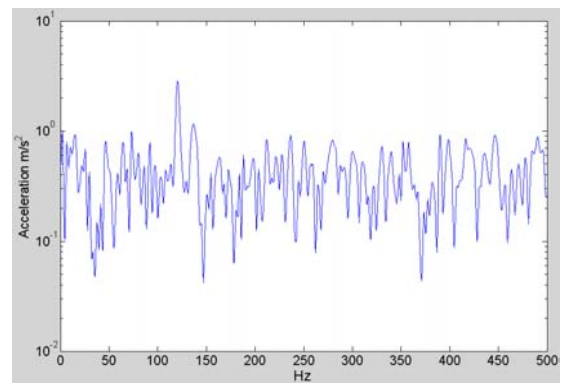

Figure 1 Typical vibration spectrum of a household appliance

## 1. INTRODUCTION

Due to the advance of CMOS VLSI technology, the power consumption of electronic devices has been reduced considerably. The low power technology enables the development of such applications as wireless sensor networks [1] or personal health monitoring [2], where remote or independent power supply is critical for building more compact or longer-life-time systems. In particular, energy scavenging from ambient natural sources, such as vibration [3], radioisotope [4] and ambient heat [5], is attracting many recent interests as the self-sustainable power source for these applications. Among various approaches, electrostatic vibration-to-electricity conversion using the micro-electro- mechanical systems (MEMS) technology is chosen in this study due to its compatibility to IC processes and the ubiquity of the energy source in nature.

The output power of a vibration driven converter is related to the nature of the vibration source, which must be known in order to estimate the generated power. The vibration spectra of several household appliances were measured. A typical vibration source has a peak acceleration of about 2.25 m/s$^2$ at about 120 Hz, as shown in Fig. 1. These values are used in the following static and dynamic analysis for the design of the converter.

## 2. DESIGN

A variable capacitor $C_v$ formed by an in-plane gap-closing comb structure is the main component in the energy converter [3, 6], as shown in Fig. 2. Fig. 3 shows a schematic circuit that can be used to extract the converted energy. The variable capacitor $C_v$ is charged by an external voltage source $V_{in}$ through the switch SW1 when $C_v$ is at its maximum $C_{max}$. When $C_v$ is charged to $V_{in}$, SW1 is opened and then the capacitance is changed form $C_{max}$ to $C_{min}$ due to the electrode displacement caused by vibration. In this process, the charge Q on the capacitor remains constant (SW1 and SW2 both open). Therefore, the terminal voltage on the capacitor is increased and the vibration energy is converted to the electrostatic energy stored in the capacitor. When the capacitance reaches $C_{min}$ ($V_{max}$), SW2 is closed and the charge on $C_v$ is transferred to a storage capacitor $C_{stor}$. SW2 is then opened and $C_v$ goes back to $C_{max}$, completing one conversion cycle. During the period when SW2 is open, the charge on $C_{stor}$ is discharged by the load resistance $R_L$ with a time constant $\tau = R_L C_{stor}$ before it is charged again by $C_v$. In the steady state, the initial and final terminal voltages $V_L$ of the discharge process become constant, as shown in Fig. 4.

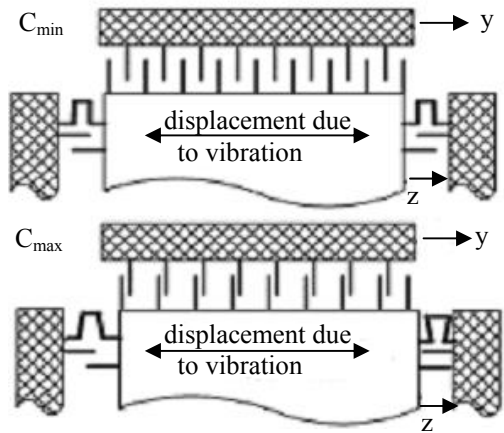

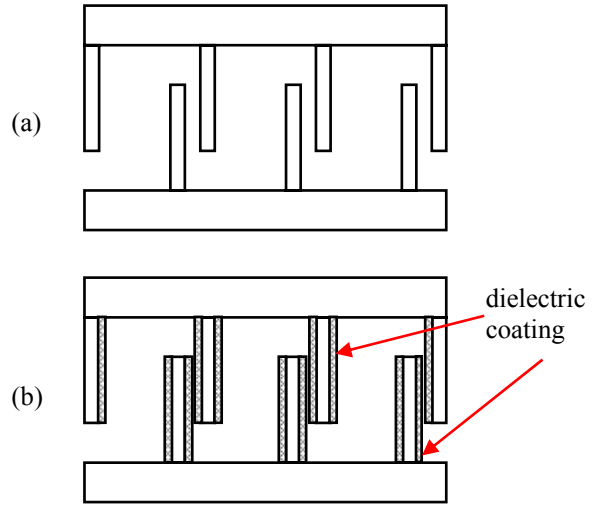

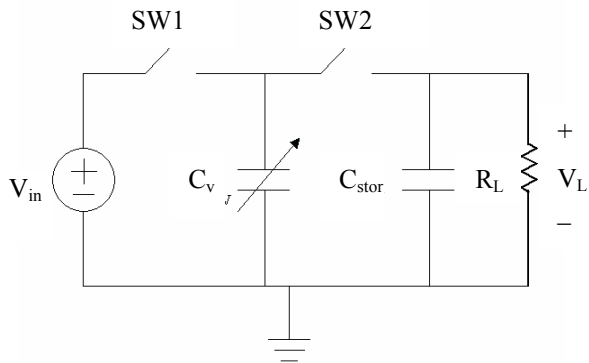

Figure 2 Variable capacitor schematic

Figure 5 Variable capacitor at $C_{max}$ position: (a) without coating, (b) with dielectric coating

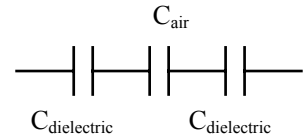

Figure 6 Equivalent $C_{total}$

Figure 3 Operation of the electrostatic energy converter

$$P_{out} = \frac{V_{sat}^2}{R_L}, \qquad (2)$$

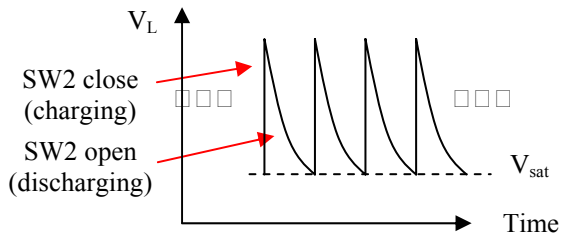

Figure 4 Output terminal voltage $V_L$ in the charge-discharge cycle

It can be shown that the steady-state final terminal voltage $V_{sat}$ in the charge-discharge cycle can be expressed as

$$V_{sat} = \frac{\frac{C_{max}}{C_{stor}} V_{in}}{\left(1 + \frac{C_{min}}{C_{stor}}\right) \times \exp(\Delta t / R_L C_{stor}) - 1}, \qquad (1)$$

where $\Delta t$ = conversion cycle time = 1/2f and f is the vibration frequency. When the voltage ripple of the charge-discharge cycle is small, as will be shown subsequently, the output power can be estimated by which is in general proportional to $C_{max}^2$. In the comb structure, $C_{max}$ is determined by the minimum finger spacing. In a previous design [7], the minimum finger spacing is kept at 0.5 μm to prevent shortage of the uninsulated fingers (Fig. 5(a)). If a dielectric coating can be applied to the side walls of the fingers (Fig. 5(b)), they become insulated and the minimum spacing can be further reduced to increase $C_{max}$ and $P_{out}$. In this design, the total capacitance becomes $C_{dielectric} \| C_{air} \| C_{dielectric}$ (Fig. 6).

Silicon nitride will be used as the dielectric material due to its process compatibility and high dielectric constant ($\varepsilon_r \sim 7$). With a 500-Å-thick nitride coating, $C_{max}$ can be increased by a factor of four, compared to the previous design. It should be noted that the dielectric coating barely increases $C_{min}$. Therefore, the expected increase of output power will not be affected by the change of $C_{min}$.

**2.1. Static analysis**

In Eq. (1), $R_L$ and $C_{stor}$ can be chosen so that the discharge time constant $\tau = R_L C_{stor}$ is much larger than the conversion cycle time $\Delta t$. The output voltage ripple in the

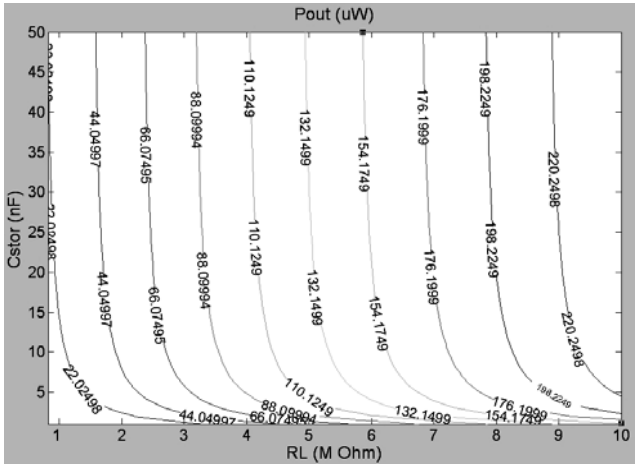

Figure 7 Output power for various $R_L$ and $C_{stor}$

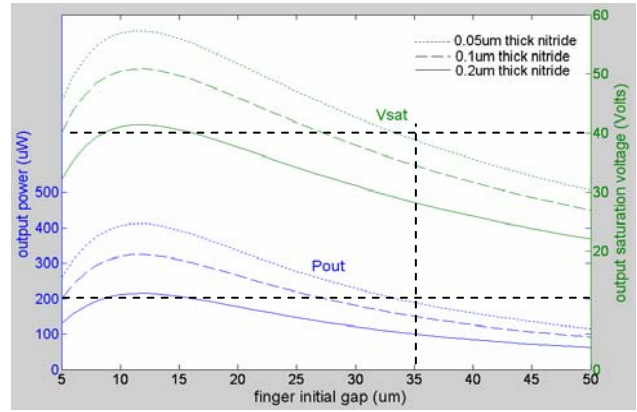

Figure 8 Output saturation voltage and power vs. initial finger gap ($R_L$ = 8 MΩ, $C_{stor}$ = 20 nF)

steady state can therefore be neglected. In this case, $V_{sat}$ can be approximated as

$$V_{sat} \approx \frac{C_{max} V_{in}}{C_{min}\left(1 + \frac{\Delta t}{R_L C_{min}} + \frac{\Delta t}{R_L C_{stor}}\right)}. \quad (3)$$

Usually $C_{min}$ is a small value (in the order of 100 pF). The other circuit components in Eq. (3) can then be chosen so that $C_{stor} \gg C_{min}$ and $R_L C_{min} \ll \Delta t$ and the expression can be simplified as

$$V_{sat} \approx \frac{C_{max} V_{in}}{C_{min} \frac{\Delta t}{R_L C_{min}}} = \frac{C_{max} V_{in}}{\Delta t} R_L. \quad (4)$$

The power output becomes

$$P_{out} \approx \frac{V_{sat}^2}{R_L} \approx \left(\frac{C_{max} V_{in}}{\Delta t}\right)^2 R_L. \quad (5)$$

For a typical low-power sensor node or module, the minimum output power requirement is about 200 μW. In addition, a power management circuit is needed to convert the high output voltage to lower ones for various sensor and signal processing units. To be compatible with the power management circuit, the maximum output voltage should be limited to about 40 V. Inserting these constraints into Eq. (2), one can obtain the range of $R_L$,

$$R_L \leq 8\ \text{M}\Omega.$$

Even though a smaller $R_L$ can be used, this would require increasing $C_{max}$ in order to satisfy the voltage and power requirement (Eqs. (4) and (5)), which in turn will have adverse effects in the dynamic behavior of the converter. Therefore, $R_L$ = 8 MΩ and hence $C_{max}$ = 7 nF are used in the following calculation.

The output power $P_{out}$ for various $C_{stor}$ and $R_L$ is shown in Fig. 7 for $C_{stor} \gg C_{min}$. It can be seen that the output power does not depend strongly on the storage capacitor $C_{stor}$ when it is relatively large. Nevertheless, a large $C_{stor}$ will result in long initial charge time when the converter starts to work from a static status. Hence, a reasonable $C_{stor}$ of 20 nF is used.

From Eq. (1) and with the values of $C_{stor}$ and $R_L$ obtained from above, input voltage $V_{in}$ of 3.3 V, vibration frequency of 120 Hz, and chip area size of 1 cm², Fig. 8 shows the calculated output saturation voltage and power as a function of the initial finger gap distance and the thickness of the silicon nitride layer. The finger thickness, length, and width are 200 μm, 1200 μm and 10 μm, respectively [7]. The dimensions of the fingers are based on the available deep etching process capability. The minimum gap distance is assumed to be 0.1 μm, which is controlled by mechanical stops. It can be seen that with a 500-Å-thick nitride, the initial finger gap has an optimal value of 35 μm for a power output of 200 μW and output voltage of 40 V.

### 2.2. Dynamic analysis

After the dimensions of the variable capacitor are determined from the static analysis, the dynamics of the micro structure is analyzed so that the desired maximum displacement, and hence $C_{max}$, can be achieved by the target vibration source. The electro-mechanical dynamics of the converter can be modeled as a spring-damper-mass system. The dynamic equation is

$$m\ddot{z} + b_e(z) + b_m(z,\dot{z}) + kz = -m\ddot{y}, \quad (6)$$

where z is the displacement of the shuttle mass m with respect to the device frame, y is the displacement of the device frame caused by vibration, $b_m(z,\dot{z})$ is the equivalent mechanical damping representing energy loss caused mainly by the squeezed film effect, and $b_e(z)$ is the electrostaitc force acting on the MEMS structure. Notice that the mechanical damping $b_m$ is a function of both the displacement z of the shuttle mass and its velocity $\dot{z}$ [3].

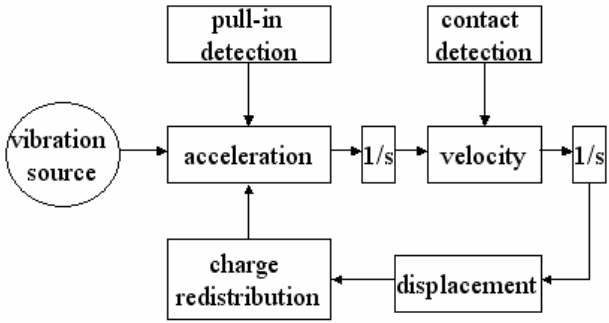

Figure 9 Dynamic simulation model

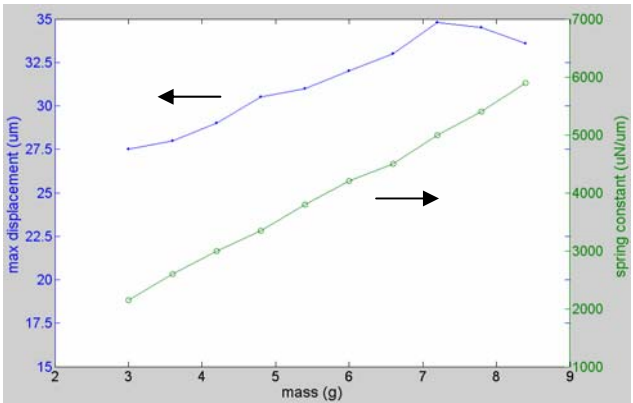

Figure 10 Maximum displacement and spring constant for various attached mass

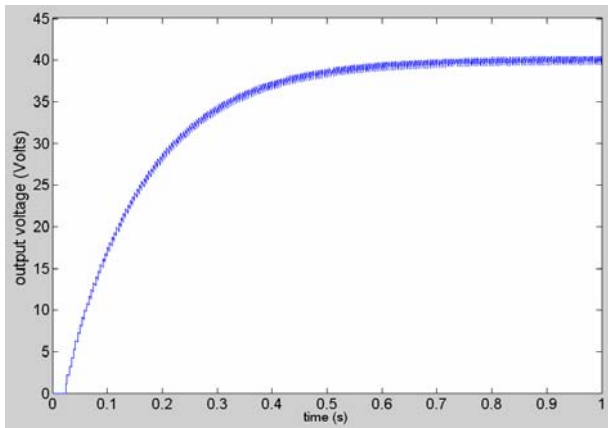

Figure 11 Output voltage vs. time

A Simulink model was built to simulate the system behavior based on Fig. 3 and Eq. (6), as shown in Fig. 9. The charge redistribution box calculates the charging and discharging events when $C_v$ reaches $C_{max}$ or $C_{min}$. This process represents the power output. Due to the limited shuttle mass that can be achieved in a MEMS process using only silicon, an external attached mass m is considered in order to increase the displacement of the variable capacitor and the energy conversion efficiency.

Table 1 Design parameters of the energy converter

| Parameter | Description | |
|---|---|---|
| W | Width of shuttle mass | 10 mm |
| L | Length of shuttle mass | 8 mm |
| $L_f$ | Length of finger | 1200 μm |
| $W_f$ | Width of finger | 10 μm |
| m | Shuttle mass | 7.2 gram |
| d | Initial finger gap | 35 μm |
| $d_{min}$ | Minimum finger gap | 0.1 μm |
| $C_{stor}$ | Storage capacitance | 20 nF |
| k | Spring constant | 4.3 kN/m |
| t | Dielectric layer thickness | 500 Å |
| $\varepsilon_r$ | Dielectric constant | 7 (SiN) |
| $R_L$ | Load resistance | 8 MΩ |
| **$V_{sat}$** | **Output voltage** | **~ 40 V** |
| **$P_{out}$** | **Output power** | **~ 200 μW** |

For various attached mass, Fig. 10 shows the maximum achievable displacement and corresponding spring constant. It can be seen that a mass of 7.2 gram is required to achieve the maximum of 34.8 μm according to the static design. The corresponding spring constant, 4.3 kN/m, will used to design the spring structures. With these values, the output voltage simulated by the Simulink model as a function of time is plotted in Fig. 11. The charge-discharge cycles are evident and the saturation voltage $V_{sat}$ is close to the expected value of 40 V. Table 1 summarizes the important device design parameters according to both the static and dynamic analysis.

## 3. FABRICATION

A schematic device layout is shown in Fig. 12. The center hole is used to fix the position of the attached steel ball. A SOI wafer with a 200-μm-thick device layer was used for large capacitance. The oxide layer and the handle wafer are 2 μm and 500 μm thick, respectively. Fig 13 shows an earlier fabrication process without the dielectric coating. The variable capacitor structure is first defined by deep reactive ion etching (Deep RIE) (Fig. 13(a)). After the sacrificial oxide layer is removed using HF solution (Fig. 13(b)), aluminum is evaporated for contact (Fig. 13(c)). A steel ball is then attached to the central plate to adjust the resonant frequency to match the vibration source and improve the conversion efficiency (Fig. 13(d)).

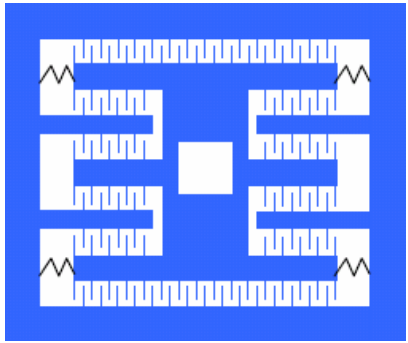

Figure 12 Layout schematic

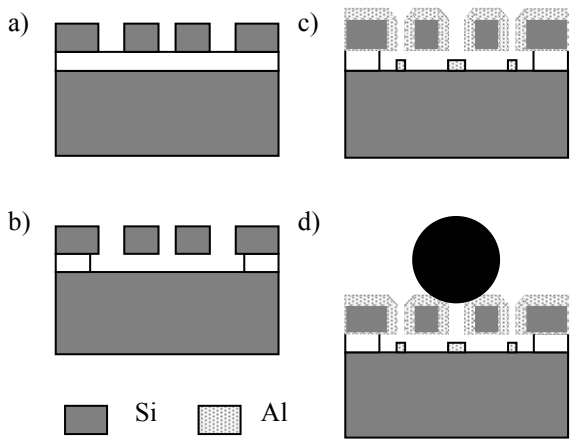

Figure 13 Fabrication process: (a) define structure by Deep RIE, (b) etch oxide by HF solution, and (c) apply Al by thermal evaporation, (d) attach external mass

The fabricated first-generation device is shown in Fig. 14 [7]. The width of the finger is reduced to 6.8 μm due to the tolerance in photolithography and RIE processes. The deviation will affect the characteristics of the converter such as the resonant frequency, output power, and output voltage.

### 4. MEASUREMENT

#### 4.1. Mechanical measurement

The displacement of the device without the attached mass was measured using a PROWAVE JZK-1 shaker. The measured response is shown in Fig. 15. Since the mass was not attached, the vibration acceleration was increased to 40 m/s$^2$ for easy observation. The maximum displacement is about 10 μm at 800 Hz, and the quality factor $Q = \omega_0/\Delta\omega$ is about 9.6, where $\omega_0$ is the resonant frequency and $\Delta\omega$ is the resonant bandwidth shown in Fig. 15. The mass of the center plate is approximately 0.038 gram, thus the spring constant can be calculated as $k = \omega_0^2 m = 960$ N/m. The measured spring constant is different from the design mainly due to the feature size shrink in the fabrication process, as shown in Fig 14.

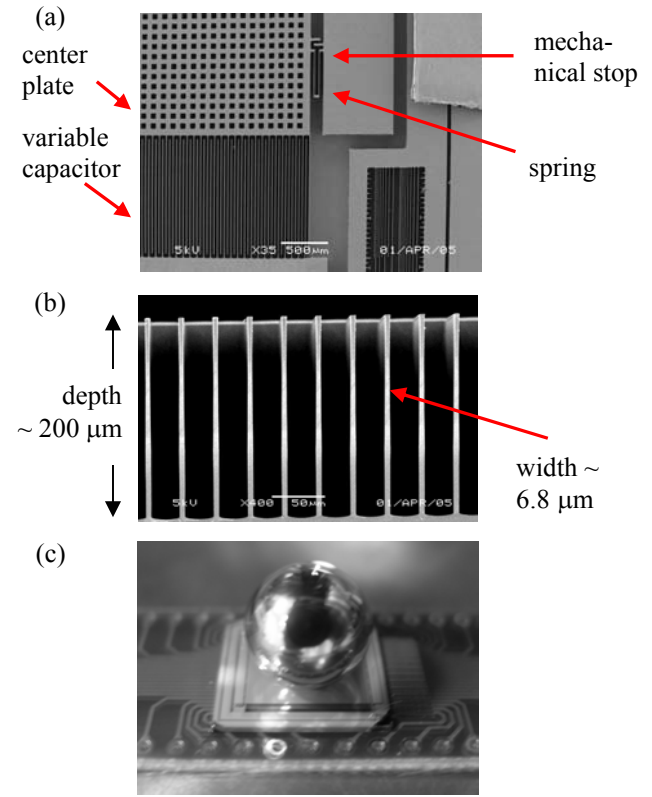

Figure 14 Fabricated device: (a) top view, (b) cross section of comb fingers, (c) overview of the converter with attached mass

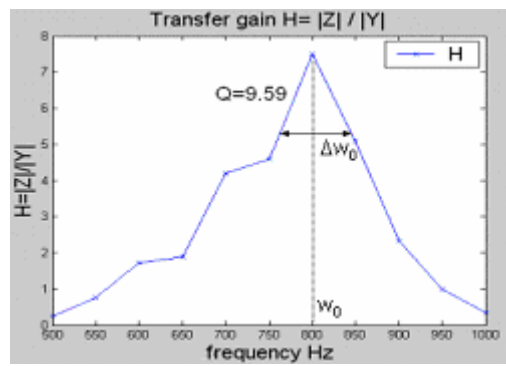

Figure 15 Frequency response of the device

#### 4.2 Electrical measurement

The electrical measurement was conducted using an INSTEK-LCR-816 LCR meter and a HP-4192A impedance analyzer. The measured capacitance without capacitance $C_{min}$ is about 50 pf. The major contribution of the large measured capacitance is the parasitic capacitance $C_{par}$ between the center plate and the substrate beneath it.

Besides the parasitic capacitance, there is also a parallel parasitic conductance. The measured conductance varies from die to die with an average resistance of 2.5 kΩ. It is suspected to be caused by the residual particles left in the device after the release step. The presence of the parasitic capacitance and conductance had hindered the measurement of output power. New devices are being fabricated with the substrate underneath the combs removed to prevent residual particles.

## 5. CONCLUSION

The design and analysis of a micro vibration-to-electricity converter are presented. The device was fabricated in a SOI wafer. The reduced feature size of the fabricated device resulted in the decrease of spring constants. Mechanical and electrical measurements of the fabricated device were conducted. Impedance measurements showed an unwanted parasitic conductance which resulted in the failure of output power measurement. Improvement of the design and fabrication processes is being conducted.

This project is supported in part by the National Science Council, Taiwan, ROC, under the grant No. NSC 93-2215-E-009-066.

## 6. REFERENCES

[1] J.M. Rabaey, et al., "Picoradio supports ad hoc ultra low-power wireless networking", IEEE Computer, Vol. 33, pp. 42-48, 2000.

[2] R. Tashiro, et al., "Development of an electrostatic generator that harnesses the motion of a living body: (use of a resonant phenomenon)", JSME International Journal Series C, Vol. 43, No. 4, pp. 916-922, 2000.

[3] S. Roundy, et al., "Micro-electrostatic vibration-to-electricity converters," Proc. IMECE 39309, 2002.

[4] R. Duggirala, et al., "Radioisotope micropower generator for CMOS self-powered sensor microsystems", Proc. PowerMEMS, pp. 133-136, 2004.

[5] T. Douseki, et al., "A batteryless wireless system uses ambient heat with a reversible-power-source compatible CMOS/SOI dc-dc converter", Proc. IEEE International Solid-State Circuits Conference, pp. 2529-33, 2003.

vibration was about 500 ~ 600 pF, while the calculated

[6] C.B. William, et al., "Analysis of a micro-electric generator for microsystems", Sensors and Actuators, A52, pp. 8-11, 1996.

[7] Y.S. Chu, et al., "A MEMS electrostatic vibration-to-electricity energy converter", Proc. PowerMEMS, pp. 49-52, 2005.